# Coding Metamaterials, Digital Metamaterials and Programming Metamaterials

Tie Jun Cui*†, Mei Qing Qi*, Xiang Wan*, Jie Zhao, and Qiang Cheng

State Key Laboratory of Millimeter Waves, Southeast University, Nanjing 210096, China

*These authors contributed equally to this work.

†To whom correspondence should be addressed. E-mail: tjcui@seu.edu.cn

As artificial structures, metamaterials are usually described by macroscopic effective medium parameters, which are named as "analog metamaterials". Here, we propose "digital metamaterials" in two steps. Firstly, we present "coding metamaterials" that are composed of only two kinds of unit cells with 0 and $\pi$ phase responses, which we name as "0" and "1" elements. By coding "0" and "1" elements with controlled sequences (i.e., 1-bit coding), we can manipulate electromagnetic (EM) waves and realize different functionalities. The concept of coding metamaterial can be extended from 1-bit coding to 2-bit or more. In 2-bit coding, four kinds of unit cells with phase responses 0, $\pi/2$, $\pi$, and $3\pi/2$ are required to mimic "00", "01", "10" and "11" elements, which have larger freedom to control EM waves. Secondly, we propose a unique metamaterial particle which has either "0" or "1" response controlled by a biased diode. Based on the particle, we present "digital metamaterials" with unit cells having either "0" or "1" state. Using the field-programmable gate array, we realize to control the digital metamaterial digitally. By programming different coding sequences, a *single* digital metamaterial has distinct abilities in manipulating EM waves, realizing the "programming metamaterials". The above concepts and physical phenomena are confirmed by numerical simulations and experiments through metasurfaces.

# 1. Introduction

Metamaterials are artificial structures that are engineered to obtain unusual properties[1], such as negative refraction[2,3], perfect lens or superlens[4], and invisibility cloaks[5-11]. The existing metamaterials are basically classified into two types based on structural arrangements: homogeneous (with periodic structures) and inhomogeneous (with non-periodic structures)[1]. The primary researches on metamaterials are periodic structures with sub-wavelength scales, which are described by uniform macroscopic medium parameters. Using special designs of unit structures, the electromagnetic (EM) waves can be controlled either by extreme values of effective permittivity and permeability (e.g. negative refraction and perfect lens owing to negative index of refraction[2-4] and high-directive emission and tunneling effect due to zero index of refraction[12-14]), or by anisotropic effective medium properties[15-17]. For the periodic structures that cannot be described by the effective medium theory, such as photonic crystals, the allowed and forbidden energy bands are used to manipulate EM waves[18,19]. On the other hand, special designs of non-periodic structures with gradiently-changed sub-wavelength geometries, which are normally behaved as macroscopic inhomogeneous metamaterials, have much more freedom in controlling EM fields. Transformation optics[5,6] provides a powerful tool in designing ideally anisotropic and inhomogeneous metamaterials to perform arbitrary controls of wave propagations, polarizations, and scattering properties to realize unusual physical phenomena and novel devices, including invisibility cloaks[5-11], concentrators[20], optical illusions[21,22], and new-concept lenses[23,24]. The conventional geometrical optics method and Fermat principle provide an alternative way to design gradient refractive index metamaterials [25] in generating high-performance antennas[26-28], high-resolution imaging lenses[29], low scattering cross sections[30], and even performing mathematical operations[31]. Recently, metasurfaces, either periodic or gradient, have also been investigated to manipulate the EM waves[32-38].

Based on the effective medium theory, existing metamaterials, either periodic or non-periodic structures, are usually characterized by continuously macroscopic media with effective permittivity and permeability, either homogeneous or inhomogeneous. The "continuous" feature makes the existing metamaterials be regarded as "analog metamaterials." Recently, Giovampaola and Engheta presented a method to construct "metamaterial bytes" by proper spatial mixtures of "digital metamaterial bits" [39], in which the "digital metamaterial bits" are some material particles with distinct material properties (e.g. Si with the positive permittivity and Ag with the negative permittivity). However, the resulted metamaterial bytes are still described by the effective medium parameters[39]. Here, we propose the general concepts of "coding metamaterial", "digital metamaterial" and "programming metamaterial". The digital metamaterial here means that a single metamaterial could be *digitally controlled* to reach distinctly different functionalities. To realize the coding metamaterial, we shall introduce two kinds of unit cells with 0 and $\pi$ phase responses to mimic "0" and "1" elements for 1-bit digital, so that they can be controlled by the existing digital technology. Such "0" and "1" elements are unnecessary to be described by macroscopic medium parameters. By designing the coding sequences of "0" and "1" elements in coding metamaterials, we can manipulate EM waves to reach different functionalities. This concept can be extended to 2-bits or more.

In fact, the opposite-phase methods have been widely used in the antenna and optical communities. Through designing a thin artificially perfect magnetic conductor (PMC, with zero phase) and combining the PMC cells with perfect electric conductor (PEC, with $\pi$ phase) cells in a *chessboard* like configuration[40], the reflections of any normally incident EM plane waves will cancel out, resulting in the reduction of radar cross sections (RCSs). Similarly, the binary-phase (or multi-phase) grating has been intensively studied in the diffractive optics, which are generated by multi-layered dielectric masks with designed etching patterns[41,42]. For example, for a phase shift of $\pi$, the etching depth has to be approximately equal to the wavelength of light[41], which is electrically large for microwave. The related area also includes spatial light modulators[43], which imposes certain forms of spatially varying modulations on the light beams. Combining with metamaterial particles, metamaterial absorber spatial light modulators were realized in the terahertz frequencies[44,45].

In this article, we propose coding metamaterials and show the abilities to manipulate EM waves by using different coding sequences of "0" and "1" metamaterial particles. We further propose a *unique* metamaterial particle that can realize either "0" or "1" element controlled by a biased diode. Then we build up a digital metamaterial, which is composed of the unique particles with either "0" or "1" state. Using the field programmable gate array (FPGA) hardware, we finally realize to control the digital metamaterial digitally. Through numerical simulation and experiments, we illustrate that a *single* digital metamaterial has distinct abilities in manipulating EM waves controlled by FPGA program, and hence realizing the "programming metamaterial".

## 2. Coding Metamaterials

We start by 1-bit coding metamaterials. As shown in Fig. 1a, we consider a special metasurface, which is composed of binary digital elements "0" or "1". The physical realization of digital elements is not unique, but requires distinct responses to reach significant phase changes, to have large freedom to control EM waves. In the binary case, the maximum phase difference is $\pi$ (or 180°). Hence we design "0" element as a metamaterial particle with 0 phase response, while "1" element with $\pi$ phase response. In this way, the phase responses of "0" and "1" elements are simply defined as: $\varphi_n = n\pi$, ($n = 0, 1$). The simplest "0" and "1" elements can be chosen as perfectly magnetic and electric conductors. However, in order to reach a broad frequency band, we make use of subwavelength square metallic patch printed on dielectric substrate to realize the binary elements (see the inset of Fig. 1b). The substrate has a thickness of $h = 1.964$ mm with dielectric constant 2.65 and loss tangent 0.001; the metallic patch has a thickness of $t = 0.018$ mm with width $w$; and the periodicity of unit cell is $a = 5$ mm. When the patch width is designed as 4.8 and 3.75 mm, the phase difference is around 180° in a broad band. In particular, from 8.1 GHz to 12.7 GHz, the phase difference ranges from 135° to 200° (is exactly 180° at 8.7 and 11.5 GHz). Hence we use the patch particle with $w=4.8$mm as "0" element, and that with $w=3.75$ mm as "1" element, which are easily fabricated in a single-layered dielectric board to construct metasurfaces. We remark that the absolute phase response of "0" element may not be 0 at a specific frequency, but it does not affect any physics since the phase can be normalized to zero.

Different from the existing analog metamaterials which make use of effective medium parameters or special dispersion relations to control EM fields, the coding metamaterials simply manipulate EM waves using different coding sequences of "0" and "1" elements. For example, under the periodic coding sequence of 010101.../010101..., the normally incident beam will be mainly reflected to two symmetrically-oriented directions by the metasurface; while under the periodic coding sequence of 010101.../101010.../010101.../101010...[40], the normally incident beam will be mainly reflected to four symmetrically-oriented directions, as illustrated in Figs. 1c and 1d. To show the above physical phenomena quantitatively, we consider a general square metasurface, which contains $N \times N$ equal-sized lattices with dimension $D$, and each lattice is occupied by a sub-array of "0" or "1" elements, as shown in Supplementary Fig. S1 in Appendix. The distribution of "0" and "1" lattices can be arbitrary. The scattering phase of each lattice is assumed to be $\varphi(m, n)$, which is either 0 or 180°. Under the normal incidence of plane waves, the far-field function scattered by the metasurface is expressed as:

$$f(\theta,\varphi) = f_e(\theta,\varphi) \sum_{m=1}^{N} \sum_{n=1}^{N} e^{-i\{\varphi(m,n)+kD\sin\theta[(m-1/2)\cos\varphi+(n-1/2)\cos\varphi]\}}, \quad (1)$$

in which $\theta$ and $\varphi$ are the elevation and azimuth angles of an arbitrary direction, and $f_e(\theta, \varphi)$ is the pattern function of a lattice. We remark that the absolute phase response of "0" element can be put into $f_e(\theta, \varphi)$ to guarantee its relative 0 phase. Hence new directivity function Dir $(\theta, \varphi)$ of the metasurface can be given by

$$\mathrm{Dir}(\theta,\varphi) = 4\pi |f(\theta,\varphi)|^2 / \int_0^{2\pi} \int_0^{\pi/2} |f(\theta,\varphi)|^2 \sin\theta d\theta d\varphi. \quad (2)$$

Here, the $f_e(\theta, \varphi)$ term has been eliminated. From above equations, we clearly observe the controls of scattered fields through coding the metasurface lattices.

For examples, when all lattices are set as "0" elements, we easily derive that $|f_1(\theta,\varphi)|=C_1|\cos\psi_1+\cos\psi_2|$; as the coding sequence is chosen as that in Fig. 1c or 1d, we have $|f_2(\theta,\varphi)|=C_2|\sin\psi_1+\sin\psi_2|$ or $|f_3(\theta, \varphi)|=C_3|\cos\psi_1-\cos\psi_2|$, in which $C_1$, $C_2$, $C_3$ are constants, and $\psi_1 = \frac{1}{2}kD(\sin\theta\cos\varphi+\sin\theta\sin\varphi)$, $\psi_2 = \frac{1}{2}kD(-\sin\theta\cos\varphi+\sin\theta\cos\varphi)$. The analyses of above expressions in Appendix show that the normally incident waves will be scattered to a single main beam, two main beams, and four main beams, respectively, using different coding patterns, which are confirmed by full-wave simulations in Fig. 2. We notice that the analytical predictions (Figs. 2d-f) have good agreements to full-wave simulations (Figs. 2g-i) in all cases. Hence we can use Eq. (1) and (2) to design complicated coding sequences in realizing advanced functionalities of coding metasurfaces.

As an application, we aim to reduce RCSs of metallic surfaces by coding "0" and "1" elements appropriately. In fact, the invisibility cloak is one approach to reduce RCSs by forcing EM waves to bend around the target[5,6], while the perfect absorber is another approach by absorbing all incident EM waves[46,47]. Here, we propose a new mechanism to reduce the mono-static and bi-static RCSs by redirecting EM energies to all directions using special "0" and "1" coding. Relative to a metallic plate with the same size, the RCS reduction caused by the coding metasurface is obtained as

$$\text{RCS reduction} = \frac{\lambda^2}{4\pi N^2 D^2} \underset{\theta,\varphi}{Max}\big(\text{Dir}(\theta,\varphi)\big). \tag{3}$$

in which $\lambda$ is the wavelength in free space. The best RCS reduction can be achieved through optimizing the coding sequences of "0" and "1" lattices, and the optimized codes for different numbers of lattices ($N$) are listed in Table I when $D$ is fixed to $\lambda$, in which the code sequences along the horizontal and vertical directions are the same. In fact, the optimized codes can operate in broadband. Although such codes are obtained at the fixed $D=\lambda$, the RCS reduction remains nearly invariant when $D/\lambda$ changes from 0.6 to 3.0, as shown in Fig. S2a. We notice that better RCS reduction is achieved for larger $N$. When $N=20$, the RCS reduction is down to -23 dB in a wide frequency band. To further guarantee the broadband performance, we show that the optimized codes are approximately valid when the phase difference between "0" and "1" elements is apart from 180° (see Fig. S2b). For all cases, when the phase difference varies from 145° to 215°, at least 10-dB RCS reduction is guaranteed.

To verify the above physical phenomena, we design and fabricate a metasurface based on the optimized coding sequences, as shown in Figs. 3a and b, which contains 8×8 lattices. The edge length of the metasurface is 280 mm, the width of each lattice is 35 mm, and each lattice is composed of 7×7 "0" or "1" elements. The patterns in Figs. 3a and b are designed symmetrically from the optimized coding sequence 00110101 for $N=8$ in both horizontal and vertical directions. In the fabrication, a commercial dielectric board (F4B) is used, which has exactly the same parameters as those in simulations. Using commercial software, CST Microwave Studio, simulation results of mono-static RCS reductions are illustrated in Fig. 3c in a wide frequency range under the normal incidence (dashed line). We notice that the 10dB bandwidth of RCS reduction in the backward direction ranges from 7.8 to 12 GHz, which is well consistent to that for phase difference. The experimental results (solid line in Fig. 3c) in a measurement system shown in Fig.S3 confirm the significant reduction of mono-static RCSs in broadband. To observe bi-static scattering features, three-dimensional (3D) scattering patterns of the metasurface are illustrated in Figs. 3d-f at three representative frequencies 8, 10, and 11.5 GHz. It is observed that the scattered fields are suppressed in low levels in all directions and the normalized bistatic RCSs are always below -10dB. This is because the coding metasurface has been designed to redirect the incident EM energies to all directions, and in each direction the energy is small based on the energy conservation principle. At 10 GHz, the small peak (below -10 dB) in the direction of incident wave is caused by the relatively large phase difference (about 203°), see Fig. 1b.

Table I. The optimized codes for different lattice numbers $N$

| $N$ | Code Sequence | RCS Reduction (dB) |
|---|---|---|
| 6 | 001011 | -12.08 |
| 7 | 0011010 | -14.64 |
| 8 | 00110101 | -15.82 |
| 10 | 0001010110 | -18.39 |
| 12 | 001001110101 | -19.75 |
| 14 | 00111110110101 | -21.41 |
| 16 | 0011110110101010 | -22.37 |
| 20 | 01000100110000110101 | -23.58 |

The concept of coding metamaterial can be extended from 1-bit coding to 2-bit or more. In 2-bit coding, four kinds of unit cells with distinct responses are required to mimic "00", "01", "10" and "11" elements, which have larger freedom to manipulate EM waves but need more complicated technology for digital controls. Similar to the 1-bit case, the four kinds of unit cells in 2-bit coding metamaterials should have phase responses of 0, $\pi/2$, $\pi$, and $3\pi/2$, corresponding to "00", "01", "10", and "11" elements. Hence the phase responses $\varphi_n$ are simply defined as: $\varphi_n=n\pi/2$, ($n$ =0, 1, 2, 3). In realization, we still make use of square metallic patches with different sizes printed on the dielectric substrate to design "00", "01", "10", and "11" elements in broadband, as shown in Fig. 4. With the 2-bit coding metamaterials or metasurfaces, we have more flexibility in controlling coding sequences to reach wider applications. For example, we design a simply periodic coding sequence 0001101100011011…, as depicted in Fig. 5a. Since the adjacent four elements "00", "01", "10", and "11" have gradient phase change, based on generalized Snell's law [32,34], the normally incident waves will be reflected to oblique angle, as confirmed by simulation results of near fields and 3D far-field scattering patterns illustrated in Figs. 5a and b. As applications, we optimize the 2-bit codes on the same-size metasurface as that in 1-bit coding to reduce RCSs of metallic surface. The optimized 2-bit codes for this purpose are displayed in Fig. 5c. Simulation results of mono-static RCSs demonstrate better performance of the 2-bit coding metasurface (see Fig. 5d), in which the RCS reduction is below -10dB in a much wider frequency band from 7.5 to 15 GHz. The 3D scattering patterns at 8, 10, 13, and 15 GHz shown in Figs. 5e-h further prove the strong ability of 2-bit coding metasurface in suppressing the bi-static RCSs.

## 3. Digital and Programming Metamaterials

In above discussions, two unit cells are used to realize "0" and "1" elements in the 1-bit coding metasurfaces. To control "0" and "1" responses digitally, we propose a unique metamaterial particle, as shown in Fig. 6a. Two planar symmetrical metallic structures (see Fig. S4) are printed on the top surface of F4B substrate with dielectric constant 2.65 and loss tangent 0.001, which are connected by a biased diode. Two metallic via holes are drilled to connect metamaterial structures with two pieces of ground, which are used to input the biased DC voltage. The total size of particle is $6\times6\times2$ mm$^3$, which is about $0.172\times0.172\times0.057\lambda^3$ at the central frequency. The biased diode can be controlled by DC voltage. When the biased voltage is 3.3V, the diode is "ON" and the corresponding effective circuit is illustrated in Fig. S5a; when there is no biased voltage, the diode is "OFF" and the corresponding circuit model is given in Fig. S5b. Inserting the circuit models into CST Microwave Studio, numerical results demonstrate the metamaterial particle to behave as "1" element with the diode on and "0" with the diode off. From Fig. 6b, we clearly notice that the phase difference is around 180° in the frequency band from 8.3 to 8.9 GHz. At 8.6 GHz, the phase difference is exactly 180°.

Based on the metamaterial particle, we design and fabricate a sample of 1-bit digital metasurface, as shown in Figs.S6a, c, and d. The digital metasurface contains $30\times30$ identical unit cells, and each includes a biased diode (see Fig. S6d). Every five adjacent columns of unit cells share a control voltage, which corresponds to a single bit of the control words. Hence this is a one-dimensional (1D) digital metasurface, and the coding sequence has 6 control words. As examples, we choose four coding sequences to validate the concepts: 000000, 111111, 010101, and 001011. Simulation results of 3D scattering patterns from the 1D digital metasurface are presented in Figs. 7a-d correspondingly. From Figs. 7a and b, we notice that the normally incident beams are directly reflected back with the coding sequences 000000 and 111111 because they mimic perfectly electric and magnetic conductors. Under periodic coding sequence of 010101, the normally incident beam is mainly reflected to two directions by the metasurface, as illustrated in Fig. 7c. In the general coding sequence of 001011, the incident beam is scattered to multiple beams with lower RCS values, as shown in Fig. 7d.

In order to control the coding sequences digitally, we design and realize FPGA hardware[48], as displayed in Fig.S6b. Four switches are used as triggers for different controls of coding sequences. When one switch is toggled on, FPGA will output the corresponding coding sequence. As a result, we could digitally change the voltage distributions on the metasurface by toggling different triggers, which further control "ON" and "OFF" of the biased diodes, producing the required "0" and "1" states of the digital metasurface. Hence the unique metasurface has different functions controlled by the FPGA program, resulting in a programming metasurface. The flow chart of the programming metasurface triggered by FPGA is illustrated in Fig. 7i. Experiments have been conducted to verify the multiple abilities of a single metasurface to manipulate EM waves. By triggering coding sequences of 000000, 111111, 010101, and 001011 with FPGA, the measured scattering patterns of the 1D digital metasurface are presented in Figs. 7e-h. From these figures, we clearly notice the multiple functionalities, which have good agreements to numerical simulations shown in Figs. 7a-d.

## 4. Conclusions

We propose the concepts of "coding metamaterials", "digital metamaterials", and "programming metamaterials", and present the designs, realizations, and experiments. For coding metamaterials, thay are composed of only two kinds of unit cells with 0 and $\pi$ phase responses, which we name as "0" and "1" elements. By coding "0" and "1" elements with the controlled sequences (i.e., 1-bit coding), we can manipulate electromagnetic waves and realize different functionalities. The concept of coding metamaterial can be extended from 1-bit coding to 2-bit or more. In 2-bit coding, four kinds of unit cells with phase responses 0, $\pi/2$, $\pi$, and $3\pi/2$ are required to mimic "00", "01", "10" and "11" elements, which have larger freedom to control the EM waves.

We further propose a unique planar metamaterial particle in the subwavelength scale, which can realize either "0" or "1" element controlled by a biased diode. Based on the novel metamaterial particle, we design and realize a digital metamaterial, in which each element has either "0" or "1" state controlled by the biased DC voltage. We have built up FPGA hardware to control the coding sequence, which is input to the digital metamaterial. Hence the single digital metamaterial can be controlled by FPGA program to reach different abilities to manipulate the electromagnetic waves, realizing a programming metamaterial.

The proposed coding metamaterials, digital metamaterials, and programming metamaterials are very attractive in a variety of applications, such as to control the radiation beams of antennas, that is similar to phase-array antennas but using different principle with much cheaper approach, to reduce the scattering features of targets, and to realize other smart metamaterials. The proposed work can be extended to the millimeter wave and terahertz frequencies.

## Acknowledgments

This work was supported in part by the National High Tech (863) Projects (2012AA030402 and 2011AA010202), in part by the National Science Foundation of China (61138001), and in part by the 111 Project (111-2-05). TJC appreciates very much to the constructive discussions with Prof. Nader Engheta at the University of Pennsylvania.

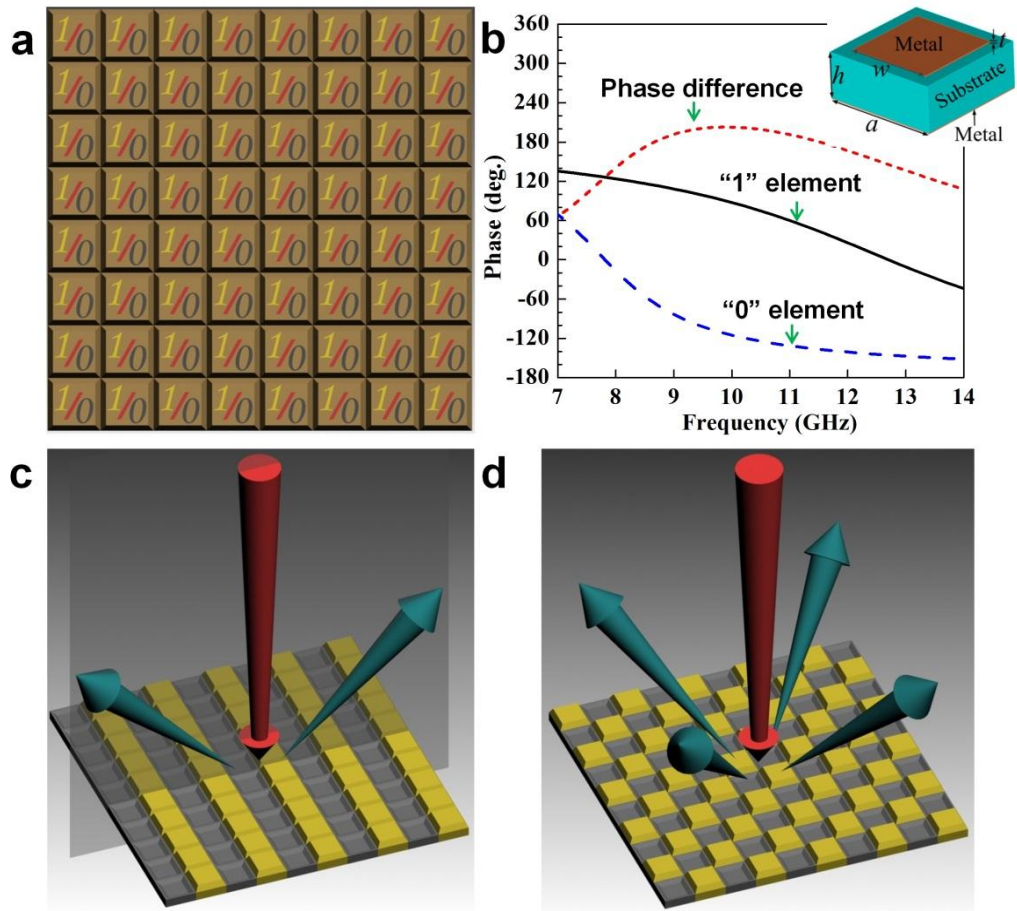

**Figure 1.** The 1-bit digital metasurface and coding metasurface. (**a**) The 1-bit digital metasurface which is composed of only two kinds of elements, "0" and "1". (**b**) A square metallic patch unit structure (inset) to realize the "0" and "1" elements and the corresponding phase responses in a range of frequencies. (**c-d**) Two 1-bit periodic coding metasurfaces to control scattering beams by designing the coding sequences of "0" and "1" elements: (**c**) the 010101…/010101... code, and (**d**) 010101…/ 101010... code.

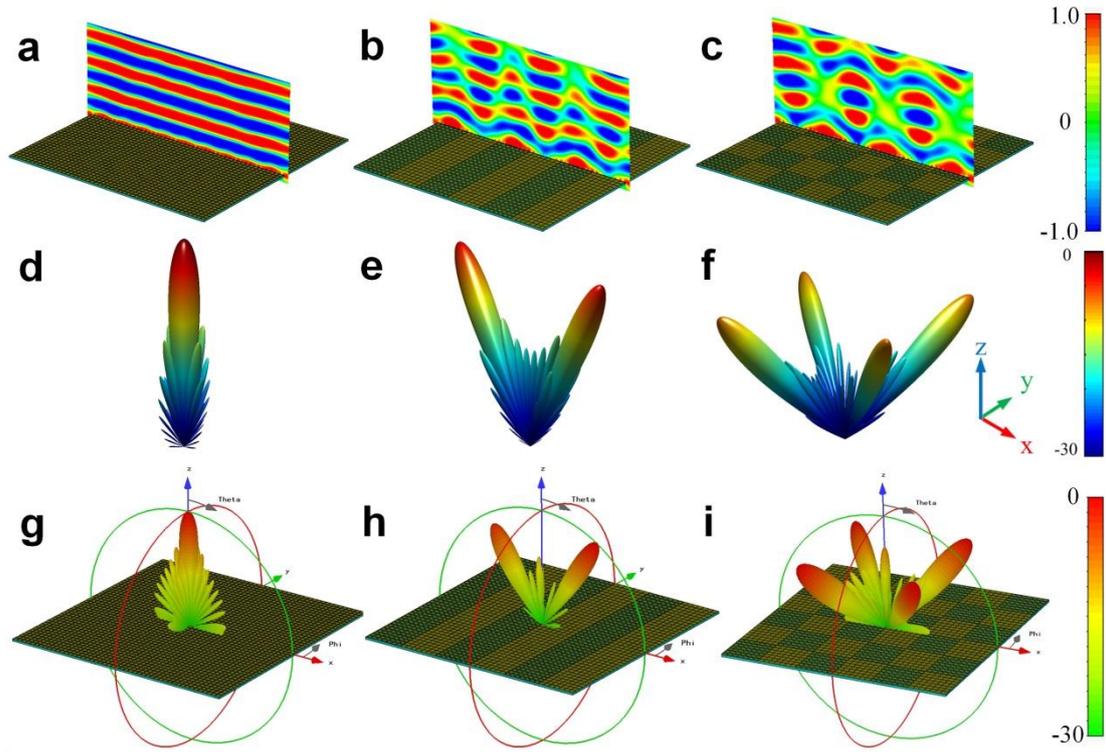

**Figure 2.** Full-wave simulation results of 1-bit periodic coding metasurfaces to show the ability in controlling scattering patterns by different coding sequences under the normal incidence of EM waves. (**a-c**) The 1-bit metasurface structures with periodic coding sequences: (**a**) 000000…/000000..., (**b**) 010101…/010101..., (**c**) 010101…/ 101010..., and their corresponding near-field distributions on the observation planes vertical to the metasurfaces. (**d-f**) The analytical results calculated by Eq. (1) of the coding metasurfaces with coding sequences: (**d**) 000000…/000000..., (**e**) 010101…/ 010101..., and (**f**) 010101…/101010..., from which a single main scattering beam, two main beams, and four main beams are generated by using different coding. (**g-i**) The full-wave simulation results of coding metasurfaces with the coding sequences: (**g**) 000000…/000000..., (**h**) 010101…/ 010101..., and (**i**) 010101…/101010..., which have good agreements to the analytical predictions.

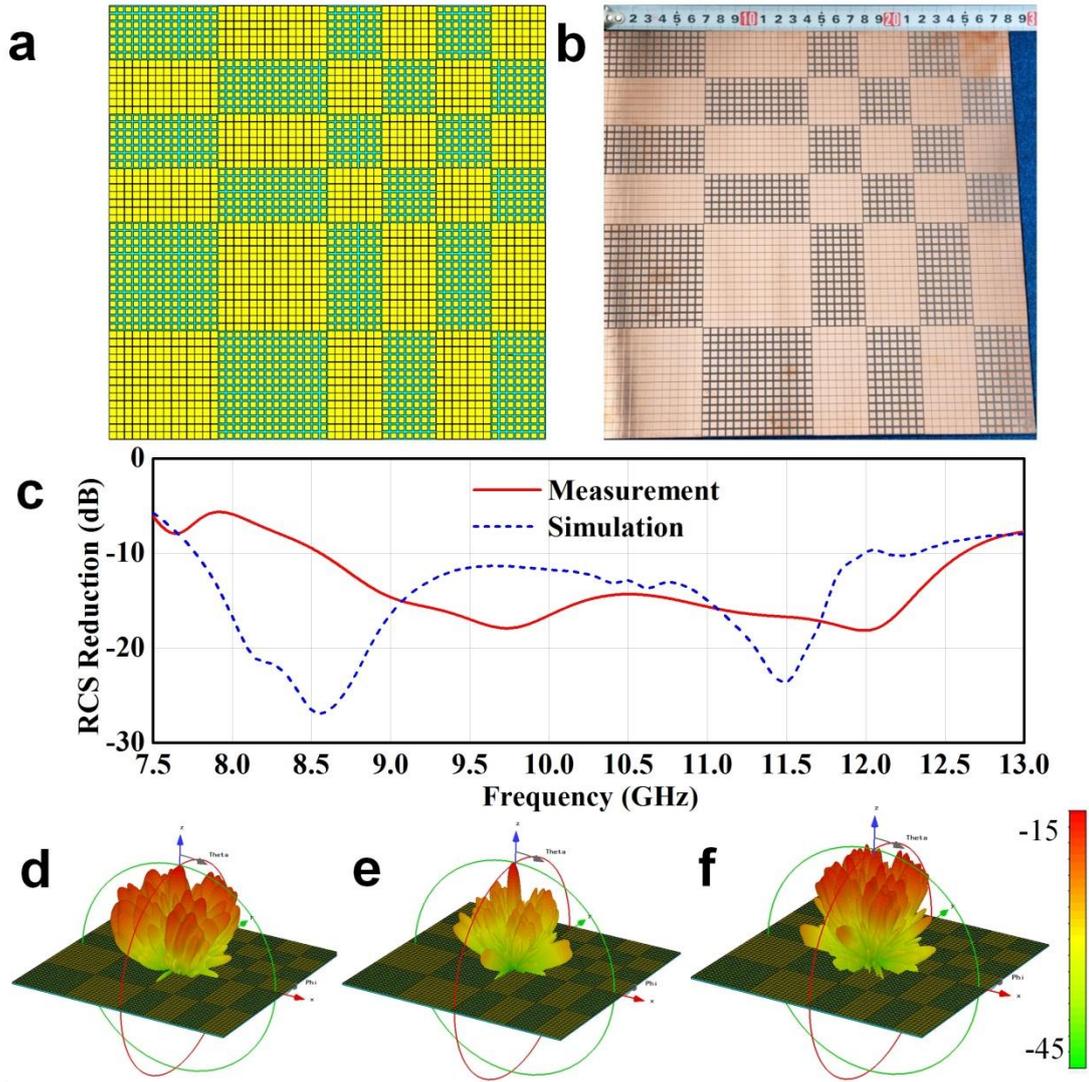

**Figure 3.** Full-wave simulation and measurement results of a non-periodic 1-bit coding metasurface to suppress the main scattering beams under the normal incidence of EM waves. (**a**) The metasurface structure with an optimized 1-bit coding sequence given in Tab. I. (**b**) The fabricated sample in F4B dielectric substrate of the optimized 1-bit coding metasurface. (**c**) The simulation and measurement results of mono-static RCS reductions in a wide frequency range from 7.5 to 13 GHz. (**d-f**) The simulation results of 3D bi-static RCS patterns at (**d**) 8 GHz, (**e**) 10 GHz, and (**f**) 11.5 GHz, respectively. It is observed that the coding metasurface redirects the incident EM energy to all directions, and the EM energy in each direction is very small, making significant reductions of both mono-static and bi-static RCSs.

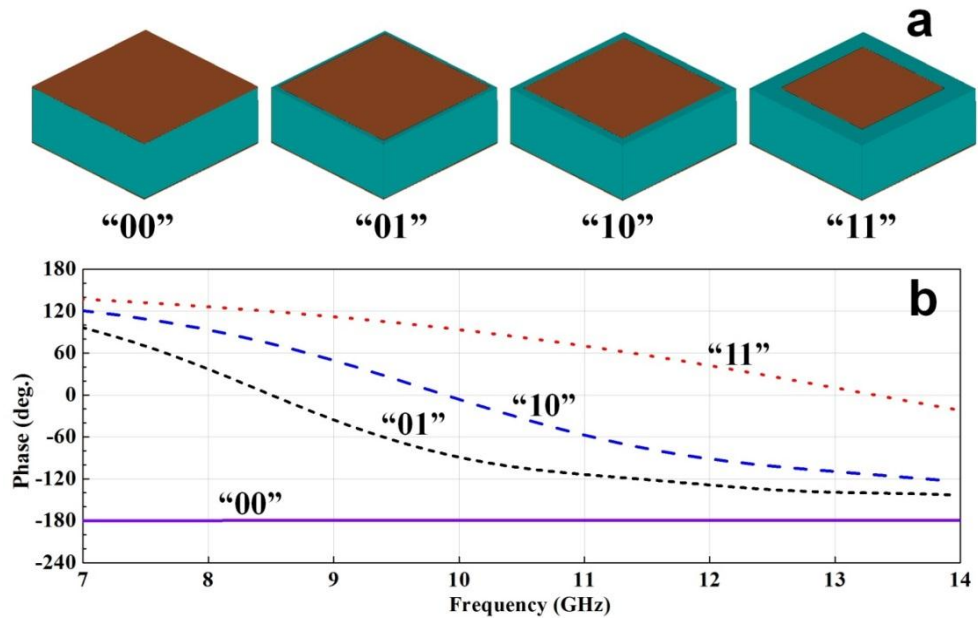

**Figure 4.** The 2-bit coding metasurface elements and their phase responses. (**a**) The "00", "01", "10", and "11" elements (from the left to the right) realized by the square metallic patches with different sizes. (**b**) The phase responses of "00", "01", "10", and "11" elements.

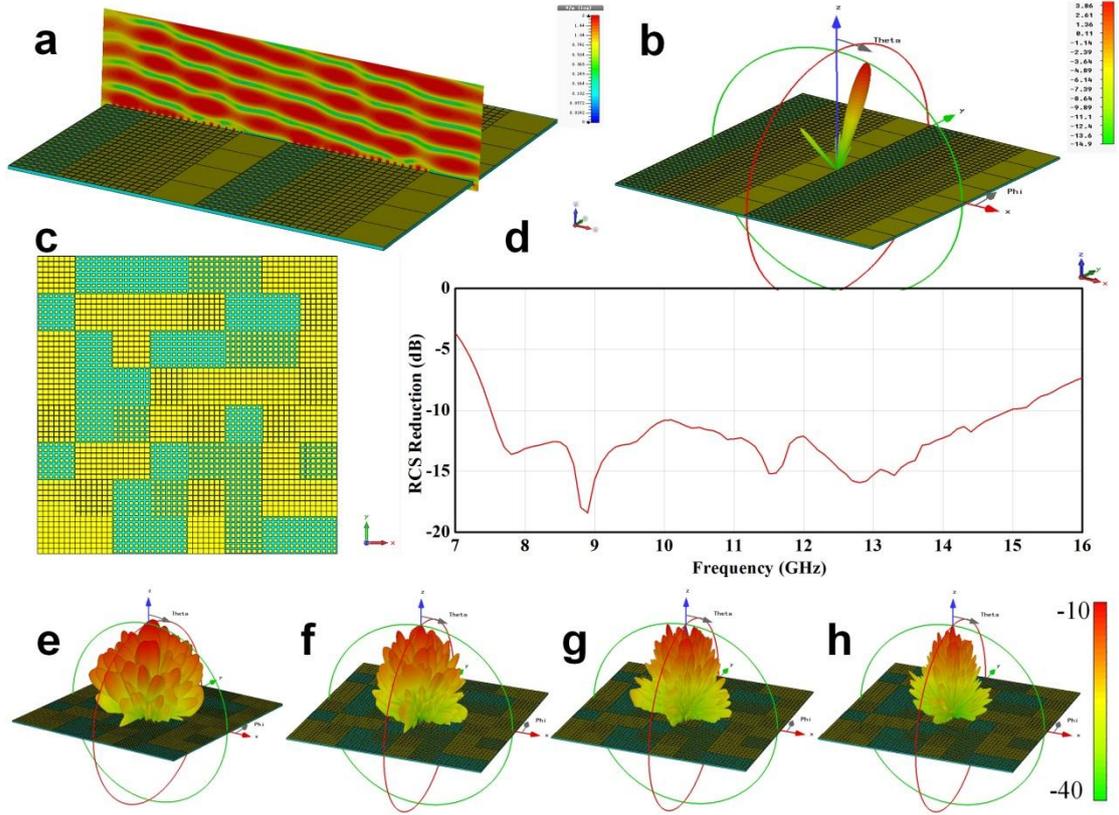

**Figure 5.** The 2-bit coding metasurfaces and their controls to EM waves under different coding sequences. (**a**) A periodic 2-bit coding metasurface with the coding sequence 0001101100011011…, and the near-field distribution on an observation plane vertical to the metasurface. (**b**) The 3D far-field scattering pattern of the periodic 2-bit coding metasurface. From near- and far-field results, we observe that the normally incident plane waves are reflected to an oblique angle, which is consistent with the generalized Snell's law due to the gradient phase change of "00", "01", "10", and "11" elements. (**c**) A non-periodic metasurface constructed by an optimized 2-bit coding sequence. (**d**) The simulation results of mono-static RCS reductions in a wide frequency range from 7 to 16 GHz. (**e-h**) The simulation results of 3D bi-static RCS patterns at 8 GHz (**e**), 10 GHz (**f**), 13 GHz (**g**), and 15 GHz (**h**), respectively. The 2-bit coding metasurface can make significant RCS reduction in a wider frequency band than the 1-bit coding metasurface.

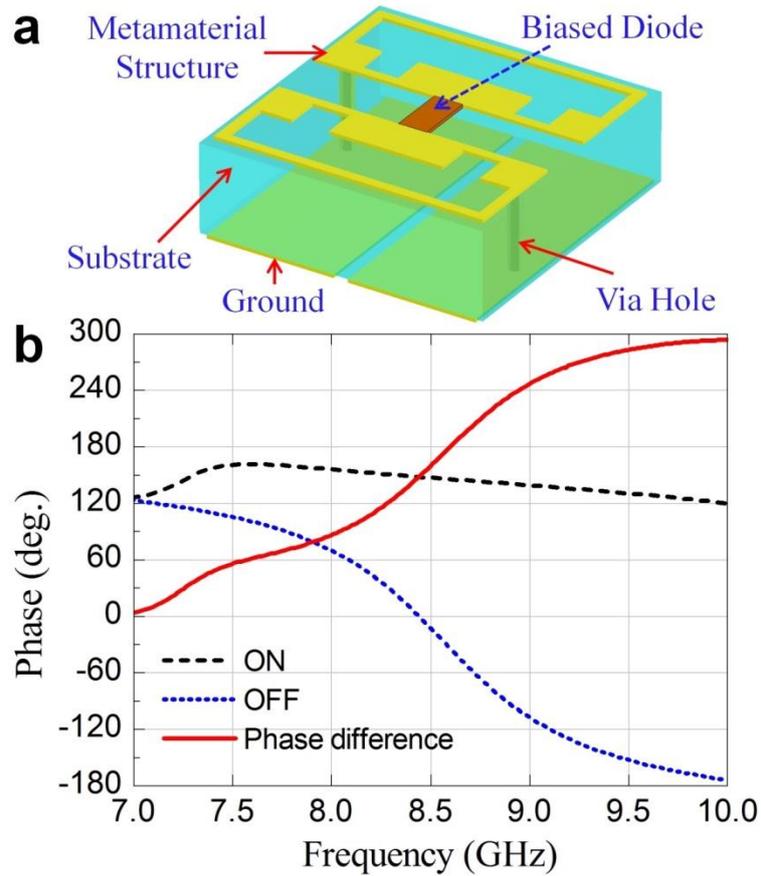

**Figure 6.** The metamaterial particle to realize digital metasurface and the corresponding phase responses. (**a**) The structure of metamaterial particle, which behaves as "0" and "1" elements when the biased diode is "OFF" and "ON". (**b**) The corresponding phase responses of the metamaterial particle as the biased diode is "OFF" and "ON" in a range of frequencies.

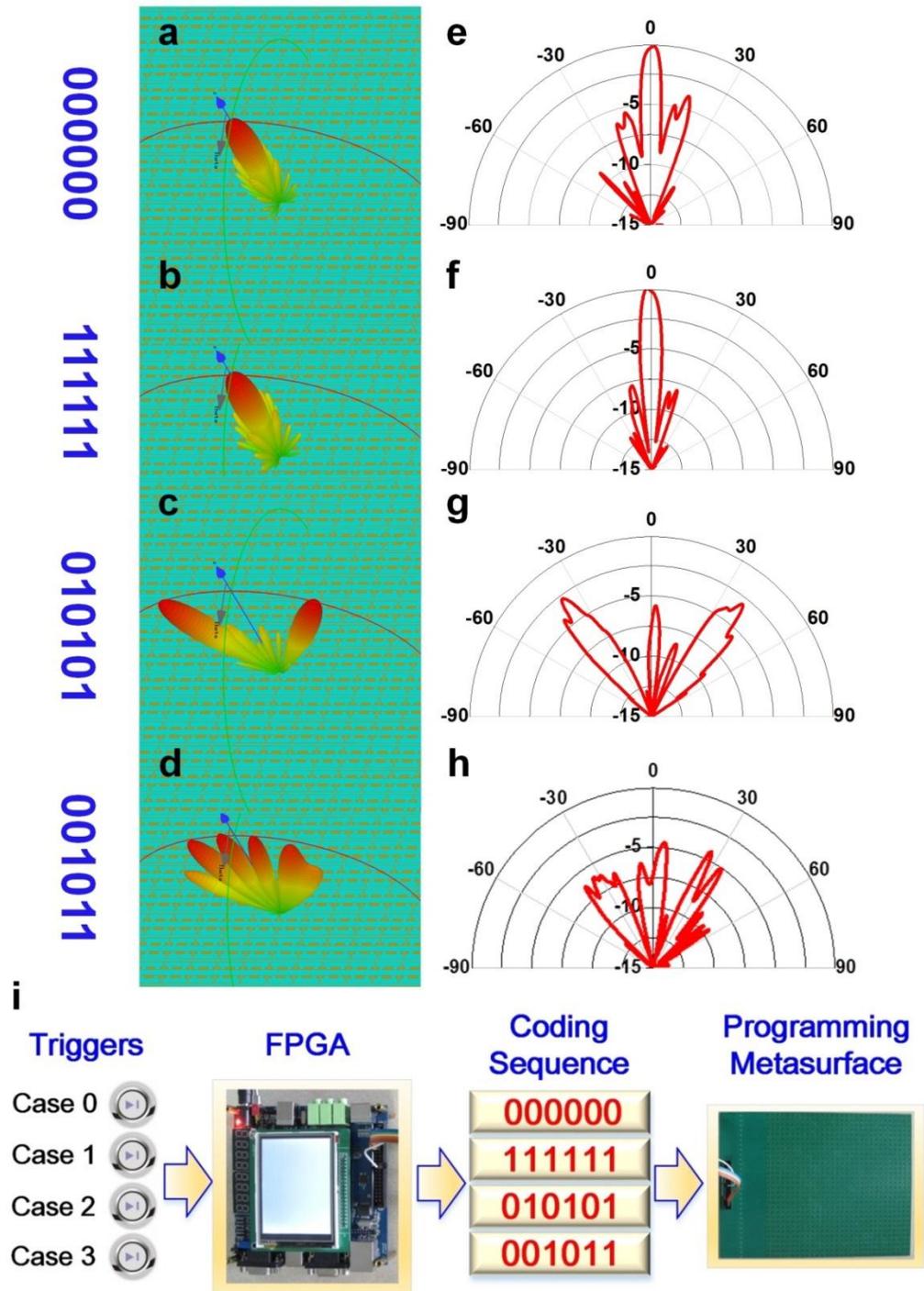

**Figure 7.** (**a-d**) Numerical simulation results of scattering patterns at 8.6 GHz for the 1D digital metasurface under different coding sequences: (**a**) 000000, (**b**) 111111, (**c**) 010101, and (**d**) 001011. (**e-f**) Experimental results of scattering patterns at 8.3 GHz for the 1D digital metasurface under different coding sequences: (**e**) 000000, (**f**) 111111, (**g**) 010101, and (**h**) 001011. (**i**) The flow diagram to realize programming metasurface controlled by the FPGA hardware.

# Appendix

## 1. Scattering Properties of 1-Bit Coding Metasurfaces

The general square metasurface is illustrated in Fig. S1, which is composed of $N \times N$ equal-sized lattices with dimension $D$. Each lattice is occupied by a sub-array of "0" or "1" elements, and the distribution of "0" and "1" lattices can be arbitrarily designed. The scattering phase of the $mn$-th lattice is assumed to be $\varphi(m, n)$, which is either 0 or 180°. Using the far-field approximation, the scattering pattern and directivity function of the metasurface are given in Eq. (1) and (2) in the main text under the normal incidence of plane waves. In the periodic coding sequences shown in Figs.2a-c, the general formula Eq. (1) can be simplified as:

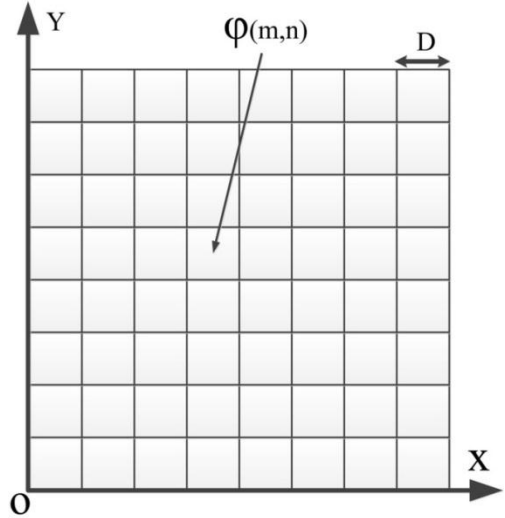

**Figure S1.** A general square metasurface, which contains $N \times N$ equal-size lattices with dimension $D$, in which each lattice is occupied by a sub-array of "0" or "1" elements.

$$|f_1(\theta,\varphi)| = C_1 |\cos\psi_1 + \cos\psi_2| = 2C_1 |\cos\frac{\psi_1+\psi_2}{2}\cos\frac{\psi_1-\psi_2}{2}|, \quad (4)$$

$$|f_2(\theta,\varphi)| = C_2 |\sin\psi_1 + \sin\psi_2| = 2C_2 |\sin\frac{\psi_1+\psi_2}{2}\cos\frac{\psi_1-\psi_2}{2}|, \quad (5)$$

$$|f_3(\theta,\varphi)| = C_3 |\cos\psi_1 - \cos\psi_2| = 2C_3 |\sin\frac{\psi_1+\psi_2}{2}\sin\frac{\psi_1-\psi_2}{2}|, \quad (6)$$

in which

$$\psi_1 = \frac{1}{2}kD(\sin\theta\cos\varphi + \sin\theta\sin\varphi), \quad \psi_2 = \frac{1}{2}kD(-\sin\theta\cos\varphi + \sin\theta\sin\varphi). \quad (7)$$

From Eqs. (4)-(6), in order to obtain the maximum scattering, the absolute values of the sinusoidal functions should be one. In the coding sequence 000000…/000000… shown in Fig. 2a, we obtain

$$|\cos\frac{\psi_1+\psi_2}{2}|=1, \quad |\cos\frac{\psi_1-\psi_2}{2}|=1. \quad (8)$$

From Eqs. (7) and (8), we easily have $\theta_1 = 0$. That is to say, the main scattering beam will be reflected to the incident indirection, which is consistent with real physics since the coding sequence 000000…/000000… represents a perfectly electric conductor with finite size. The analytical calculation and full-wave simulation results presented in Figs.2d and g confirm the conclusion.

When the coding sequence is 010101…/010101… shown in Fig. 2b, we have

$$|\sin\frac{\psi_1+\psi_2}{2}|=1, \quad |\cos\frac{\psi_1-\psi_2}{2}|=1. \quad (9)$$

From Eqs. (7) and (9), we derive that $\varphi_2$=90° and 270°, and $\theta_2 = \arcsin\frac{\lambda}{2D}$. In this case, we conclude that there are two main scattering beams directing to ($\theta_2$, 90°) and ($\theta_2$, 270°), as illustrated in Figs. 2e and h. When the coding sequence is 010101…/101010…, as depicted in Fig. 2c, we have

$$|\sin\frac{\psi_1+\psi_2}{2}|=1, \quad |\sin\frac{\psi_1-\psi_2}{2}|=1. \quad (10)$$

Similarly, we derive that $\varphi_3$=45°, 135°, 225°, 315°, and $\theta_3 = \arcsin\frac{\lambda}{\sqrt{2}D}$, which imply four scattering beams directing to ($\theta_3$, 45°), ($\theta_3$, 135°), ($\theta_3$, 225°), and ($\theta_3$, 315°), as confirmed in Figs. 2f and i.

## 2. Broadband Features of Coding Metasurfaces

Although the optimized codes in Table I are obtained when $D$ is fixed to $\lambda$, they in fact can be used in broadband for different numbers of lattices, as illustraed in Fig. S2a. From this figure, we notice that the RCS reduction remains nearly invariant when the lattice dimension $D$ changes from $0.6\lambda$ to $3.0\lambda$. To further guarantee the broadband performance, the code sequences should work for phase differences other than 180° because it is hard to practically realize "0" and "1" elements that strictly have the opposite phases over a broad frequency band. Hence we inspect the relation between RCS reduction and phase difference as a function of $N$, as shown in Fig. S2b. We observe that 10-dB RCS reduction is achieved when the phase difference varies from 145° to 215°. The RCS reduction remains nearly constant when the phase difference is around 180°. As the phase difference is far away from 180°, the RCS reduction gradually degenerates since the cancellation effect goes worse towards the incident direction. From Fig.S2b, we also notice better RCS reductions for larger $N$.

## 3. Measurement System

To measure the normalized RCSs in the specular angle, a simple experimental setup is established, which includes a transmitting (Tx) horn antenna, a receiving (Rx) horn antenna, and an Agilent Vector Network Analyzer (N5230C), see Fig. S3. The Tx and Rx horn antennas are placed in the specular directions with respect to the coding metasurface sample, which are connected to Network Analyzer to excite and receive EM signals. To eliminate the interference of environment, the function of time-domain gating in the Network Analyzer is adopted in experiments.

## 4. 2-Bit Coding Metasurfaces

The 2-bit coding metasurfaces includes four basic elements "00", "01", "10" and "11", which have relative phase responses of 0, $\pi/2$, $\pi$, and $3\pi/2$, respectively. To realize such elements, we still use the square metallic patches printed on a dielectric substrate, as shown in Fig. 4a. When the patch size is chosen as $w$=5, 4.68, 4.4, and 3.6 mm, the particle will mimic the "00", "01", "10", and "11" elements, respectively. Fig. 4b illustrates the corresponding phase responses, which satisfy the required phase shifts with a tolerance in broadband.

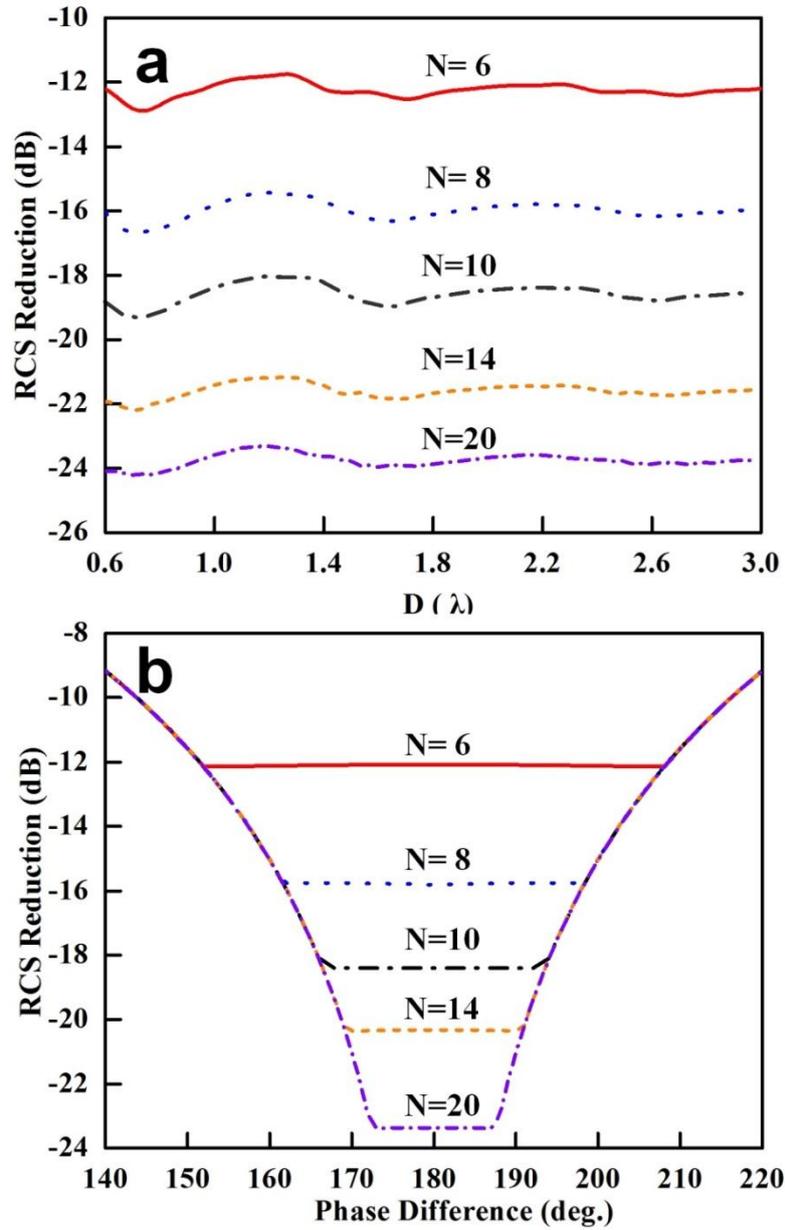

**Figure S2.** Broadband features of the optimized coding metasurfaces for RCS reduction. (**a**) The RCS reduction versus the lattice electric length $D/\lambda$ when $N$ is different, in which the phase difference is fixed as 180°. (**b**) The RCS reduction versus the phase difference of "0" and "1" elements when $N$ is different, in which $D$ is fixed to $\lambda$. In both cases, good broadband features are observed to suppress RCSs.

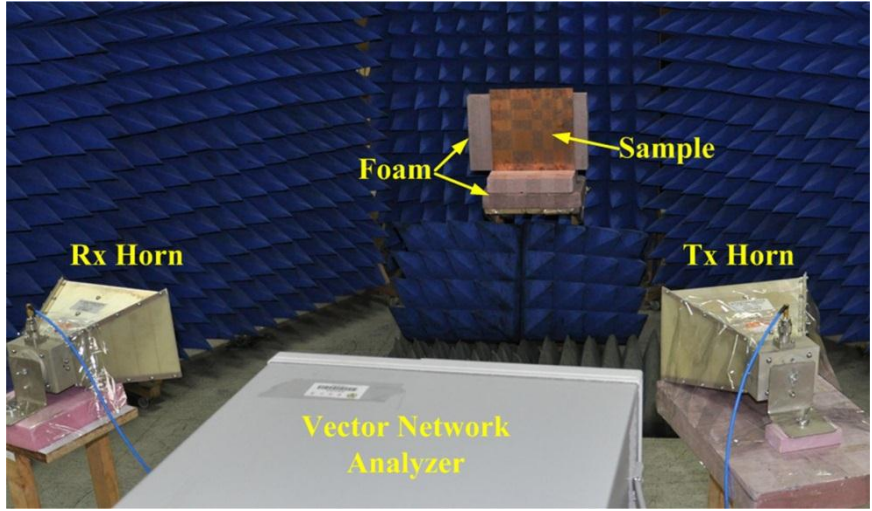

**Figure S3.** The photograph of experimental setup for RCS measurements, which is composed of a transmitting (Tx) horn antenna, a receiving (Rx) horn antenna, and an Agilent Vector Network Analyzer (N5230C).

## 5. Digital and Programming Metasurface

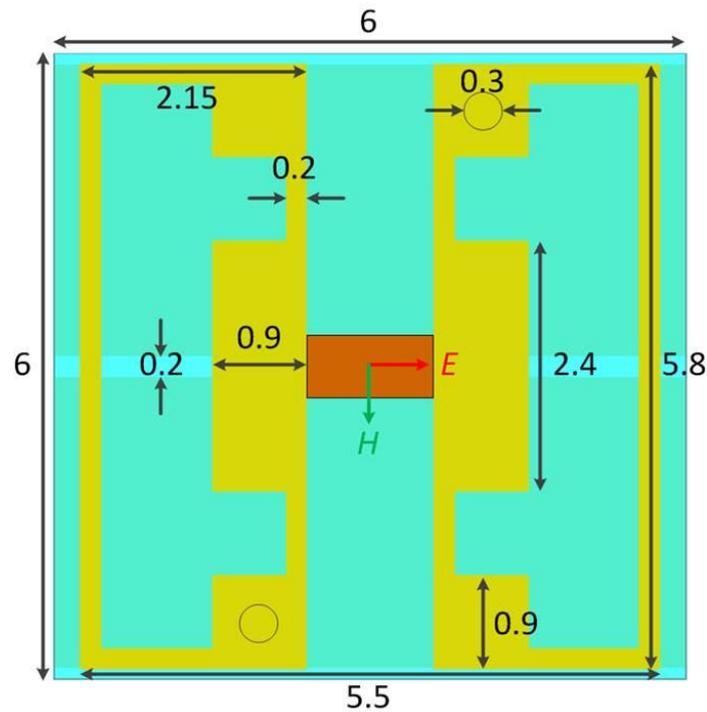

**Figure S4.** Top view of the unique metamaterial particle to realize digital metasurface, in which the detailed geometrical parameters are given.

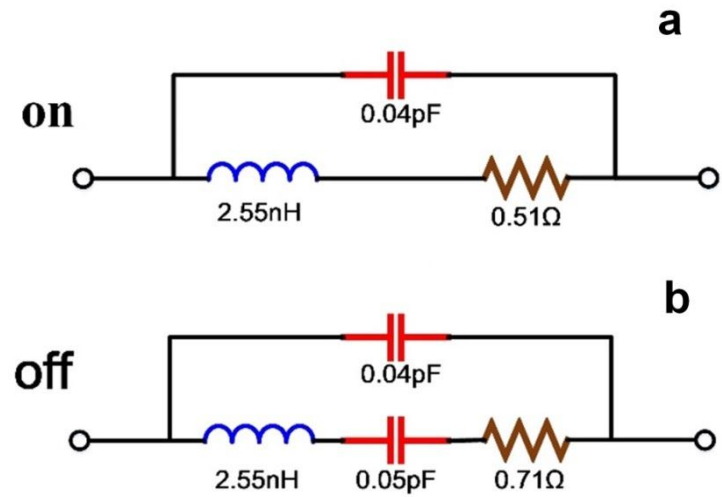

**Figure S5.** The effective circuit models of the biased diode at the "on" and "off" states. (**a**) The "on" state. (**b**) The "off" state.

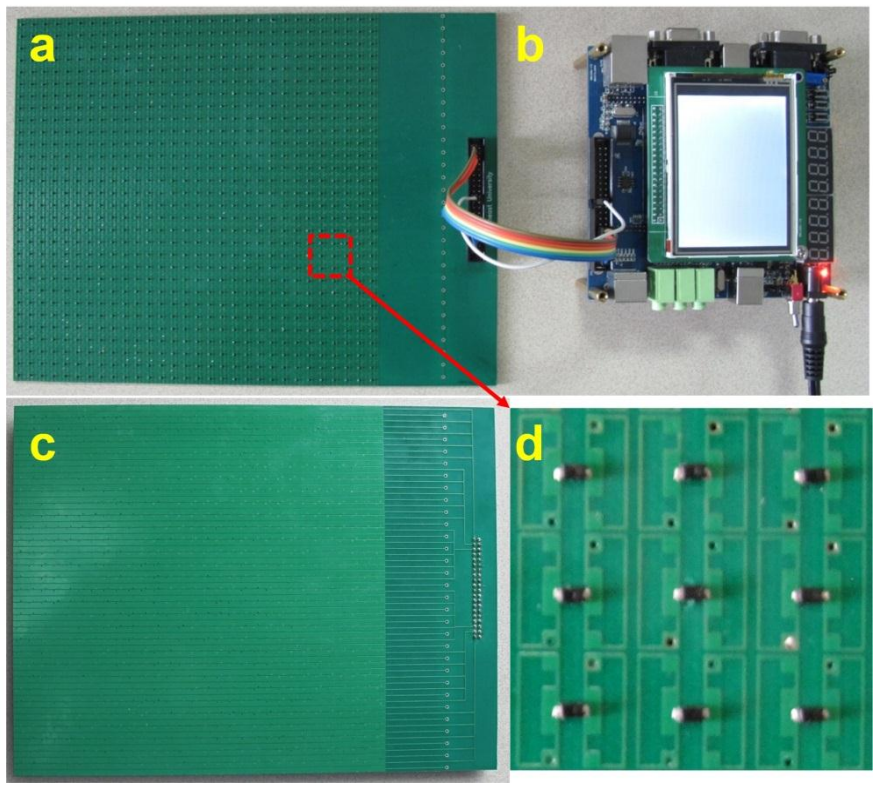

**Figure S6.** The fabricated 1D digital metasurface controlled by FPGA. (**a**) The top view of the 1D digital metasurface. (**b**) FPGA. (**c**) The bottom view of the 1D digital metasurface. (**d**) The zoomed view of the 1D digital metasurface.